\documentclass[aps,amsfonts,pra,twocolumn,superscriptaddress]{revtex4-1}
\usepackage{epsfig,amsmath,amssymb,bm,epsf,graphicx,bbold,dsfont,cancel}
\usepackage[all]{xy}
\usepackage{color}
\usepackage{physics}

\begin{document}

\title{The AKLT models on the singly decorated diamond lattice and two  degree-4 planar lattices are gapped}
\author{Wenhan Guo}
\affiliation{C. N. Yang Institute for Theoretical Physics and Department of Physics and Astronomy, State University of New York at
Stony Brook, Stony Brook, NY 11794-3840, USA}
\author{Nicholas Pomata}
\affiliation{C. N. Yang Institute for Theoretical Physics and Department of Physics and Astronomy, State University of New York at
Stony Brook, Stony Brook, NY 11794-3840, USA}
\author{Tzu-Chieh Wei}
\affiliation{C. N. Yang Institute for Theoretical Physics and Department of Physics and Astronomy, State University of New York at
Stony Brook, Stony Brook, NY 11794-3840, USA}

\begin{abstract}
Recently various 2D AKLT models have been shown to be gapped, including the one on the hexagonal lattice. Here we report on a non-trivial 3D AKLT model
which consists of spin-2 entities on the diamond lattice sites and one single spin-1 entity between every neighboring spin-2 sites.  Although  the nonzero gap problem for the uniformly spin-2 AKLT models on the diamond and square lattices is still open, we are able to establish the existence of the gap for two planar lattices, which we call the inscribed square lattice and the triangle-octagon lattice, respectively. So far, these latter two models are the only two uniformly spin-2 AKLT models that have a provable nonzero gap above the ground state.  We also discuss some attempts in proving the gap existence on both the square and kagome lattices. In addition, we show that if one can solve a finite-size problem of a weighted AKLT Hamiltonian and if the gap is larger than certain threshold, then the model on the square lattice is gapped in the thermodynamic limit. The threshold of the gap scales inversely with the linear size of the  finite-size problem. \end{abstract}

\date \today
 \maketitle

\section{Introduction}
The spin models constructed by Affleck, Kennedy, Lieb and Tasaki (AKLT) in 1987~\cite{AKLT1,AKLT2} have prompted many further developments. This includes symmetry-protected topological phases~\cite{Gu,Pollmann,Chen}, where the spin-1 chain notably exemplifies the one-dimensional Haldane phase~\cite{Haldane83} and the two-dimensional spin-2 model realizes Haldane's 2D disordered phase~\cite{Haldane83b}. One key property for such phases of matter to be stable is the existence of a nonzero energy gap above the ground state. In one dimension, this was already solved in the original AKLT spin-1 chain and general methods have been proposed and successfully applied~\cite{AKLT1,Knabe,Fannes1992}. 
Another, unexpected development in the study of AKLT states is their application to  quantum computation~\cite{Gross,Brennen,Wei11,Miyake,Wei2013,Wei2014,Wei15}. In particular, the use of certain two-dimensional AKLT states under local measurements can give rise to universal quantum computation~\cite{Wei11,Miyake,Wei2013,Wei2014,Wei15}.  The existence of a gap could be useful in order to obtain the ground state by cooling the engineered Hamiltonian.

 AKLT's original conjecture, that the spin-3/2 AKLT model on the hexagonal/honeycomb lattice indeed has a nonzero spectral gap~\cite{PomataWei2020,LemmSandvikWang2020}, has only been recently proved.  Both works employed analytical simplification of the gap criteria to numerically verify them, with high enough precision that no doubt could remain.
 The demonstration of the nonzero gap was also carried out in Ref.~\cite{PomataWei2020} on other 2D degree-3 lattices (with uniformly spin-3/2 degrees of freedom) and two singly decorated ones (spin-1 mixed with, respectively, spin-3/2 and spin-2). But no AKLT models with uniformly spin-2 degrees of freedom have yet been shown to be gapped in the thermodynamic limit.
 Here, we employ the method of Ref.~\cite{PomataWei2020} to several other  AKLT models on: (1)  the 3D singly decorated diamond lattice (of spin-2 and spin-1 mixture); see Fig.~\ref{fig:DecoratedDiamondFour}, (2) the triangle-octagon lattice; see Fig.~\ref{fig:OctagonTriangleLattice}, and (3) the 2D `inscribed square lattice'; see Fig.~\ref{fig:InscribedSquareLattice},  with the latter two being uniformly spin-2. 
 
 Ref.~\cite{Abdul} was the first to discuss the issue of the gap on decorated honeycomb lattices, and proved the existence of the gap for the number $n$ of decorations on each edge being $3$ or greater. Similar decorated lattices for the AKLT models were also discussed previously in the context  of measurement-based quantum computation in Ref.~\cite{Wei2014}.
 Extending the work of Ref.~\cite{Abdul},  Ref.~\cite{PomataWei2019} showed that the issue for multiply decorated ($n\ge 2$) lattices (for which the degree in the original ones is 6 or less) reduces to a problem involving two original lattice sites and incident edges with decoration. This can be solved without the knowledge of nearby local geometry, be the undecorated lattice  two-dimensional, three-dimensional or even higher. The unresolved problem of {\it singly} decorated lattices may require the knowledge of nearby local geometry, as demonstrated in the 2D singly decorated square lattice and hexagonal lattice~\cite{PomataWei2020}. They are the variant closest to the AKLT model on the original (undecorated) lattice. Thus, the case of the singly decorated diamond lattice provides  a non-trivial  3D AKLT  model  to examine its spectral gap issue.

The spin-2 AKLT model on the original, undecorated diamond lattice is disordered~\cite{Param}, but a sufficient condition for this to hold is the existence of a nonzero gap above the ground state, which is still evasive at the moment. The related singly-decorated model, with a single spin-1 entity added to every edge of the diamond lattice, is also likely to be disordered as
the spin-1 configuration forms the AKLT chain, which is known to be disordered due to its greater quantum fluctuations.
Here, we show that this decorated model is indeed gapped and we also provide several different approaches for lower-bounding the energy gap, one of which yields $\Delta_{\rm lower}\ge 0.013622$. The value is likely much lower than the actual gap.  This shows that the disordered property of the model is stable against small perturbations, which is a nontrivial result in a three-dimensional AKLT model. 

Additionally, we are able to establish the nonzero gap for AKLT models on two planar lattices where every site has spin 2, which we call the inscribed square lattice and the triangle-octagon lattice, respectively.  Despite these results, we are still not able to do that for the square lattice or the kagome lattice.  But we discuss a few options toward that goal. Moreover, by using the approach introduced in Ref.~\cite{LemmSandvikWang2020}, we also prove a finite-size criterion for the square lattice. We show that if one can solve a finite-size problem of a weighted AKLT Hamiltonian and if the gap is larger than a certain threshold, then the AKLT Hamiltonian on the square lattice is gapped in the thermodynamic limit. The threshold of the gap scales inversely with the linear size of the problem. This means that if numerical methods, such as DMRG or other tensor network methods, were able to demonstrate  a gap larger than the threshold for a certain size, then the issue of the AKLT gap on the square lattice would be solved. 

The remainder of the paper is organized as follows. In Sec.~\ref{sec:method}, we review the method of proving the gap and provide more detailed discussions on how to obtain the lower bound on the energy gap.  Different approaches for the lower bounds on the gap are presented. In Sec.~\ref{sec:results}, we give our results of demonstrating the nonzero gap of the AKLT models on the singly decorated diamond lattice and two other planar lattices.  In Sec.~\ref{KagomeSquare}, we describe our attempts and ideas to tackle the gap issue on the kagome and square lattices. In Sec.~\ref{Finite}, we generalize the finite-size method in Ref.~\cite{LemmSandvikWang2020} and derive a corresponding criterion for establishing the gap  in the AKLT model on the square lattice. We conclude in Sec.~\ref{sec:conclusion}.

\section{Key methods}
\label{sec:method}

Following the procedure developed in Refs. \cite{PomataWei2020,PomataWei2019}, we first decompose the graph (corresponding to the lattice in question) into overlapping subgraphs, which collectively contain all the edges. The original AKLT Hamiltonian can then be rearranged according to these subgraphs, with a suitable weight for each edge. We can use projectors supported on these subgraphs to construct a Hamiltonian that lower-bounds the original AKLT Hamiltonian. If such a lower-bounding Hamiltonian can be shown to be gapped, then the original AKLT Hamiltonian is gapped and a lower bound on the gap may also be obtained.

\begin{figure}
 {\centering
 \includegraphics[width=0.45\textwidth]{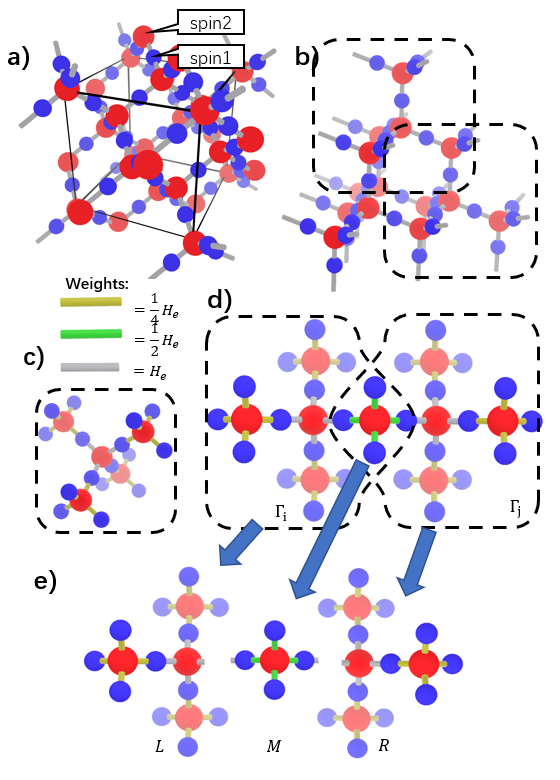}}
 \caption{\label{fig:DecoratedDiamondFour}
 (a) The decorated diamond lattice. (b) The overlapping scheme. (c) The subgraph $\Gamma$. (d) A pair of overlapping subgraphs. (e) The three parts of this pair of overlapping subgraphs. } 
 \end{figure}

\begin{figure}
 {\centering
 \includegraphics[width=0.3\textwidth]{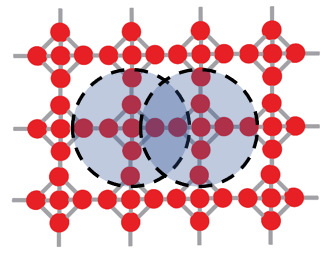}}
 \caption{\label{fig:OctagonTriangleLattice}
The triangle-octagon lattice  and the proposed overlapping scheme.
}
\end{figure}

\begin{figure}
 {\centering
 \includegraphics[width=0.28\textwidth]{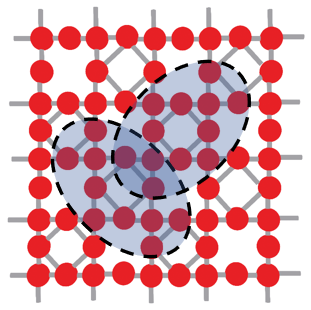}}
 \caption{\label{fig:InscribedSquareLattice}
Inscribed Square Lattice, and the overlapping scheme.
}
\end{figure}

There are multiple choices of subgraphs, but there is not a generic approach to determine which one yields a successful proof of the nonzero gap. Examples for the degree-3 Archimedean lattices were constructed in Ref.~\cite{PomataWei2020} and those for the three models solved in this work are shown in Figs.~\ref{fig:DecoratedDiamondFour},~\ref{fig:OctagonTriangleLattice} and~\ref{fig:InscribedSquareLattice}.
It is thus necessary to check explicitly whether the gap criterion is satisfied. We shall discuss this criterion below;  but, in brief, it relies on a relation between two nearby projectors supported on two overlapping subgraphs in terms of their sum and anti-commutator, can be satisfied; see Eqs.~\eqref{eq:lower} and~\eqref{eq:anticomm} below. However, even for two subgraphs with relatively small sizes, the central eigenvalue problem may have too large a dimension to solve on any computer. The tensor-network method introduced in Ref.~\cite{PomataWei2020} will be used in reducing the problem size substantially, and if it becomes solvable on a computer, the gap criterion can be checked with high precision.

\subsection{Lower-bounding the Hamiltonian using projectors}

Given a set of subgraphs $\Gamma_i$, the AKLT Hamiltonian can be rearranged as the sum of weighted AKLT Hamiltonian of the subgraphs,
\begin{equation}
    H_{AKLT}=\sum_e H_e=\sum_i\sum_{e\in\Gamma_i}w_e^{(\Gamma_i)}H_e.
\end{equation}
We then bound the subgraph AKLT Hamiltonian $H_{\Gamma_i}$ with the orthogonal projector $\tilde{H}_i$ that has the same ground space as $H_{\Gamma_i}$,
\begin{equation}
    H_{\Gamma_i}=\sum_{e\in\Gamma_i}w_e^{(\Gamma_i)}H_e\geq\gamma_0\tilde{H}_i,
\end{equation}
where  $\gamma_0$ is the least nonzero eigenvalue of $H_{\Gamma_i}$.

\subsection{Gappedness and gap lower bound}
If we can prove the lower bounding Hamiltonian $\tilde{H}=\sum_i\tilde{H}_i$ has a gap $\tilde{\gamma}$, i.e.,
\begin{equation}\label{eq:gammaH}
    \tilde{H}^2\geq\tilde{\gamma}\tilde{H},
\end{equation}
then the AKLT Hamiltonian is gapped.
To do this, we consider contributions to $\tilde{H}^2$. 
Since non-overlapping pairs of Hamiltonian terms commute,  we can  thus discard these positive semi-definite terms, yielding a lower bound on $\tilde{H}^2$, 
\begin{subequations}
\label{eq:H2}
\begin{eqnarray}
   \tilde{H}^2&=&\sum_i\tilde{H}_i+\sum_{i\neq j}\tilde{H}_i\tilde{H}_j\\
    &\geq& \tilde{H}+\sum_{<i,j>}\{\tilde{H}_i,\tilde{H}_j\}\geq(1-\tilde{z}\eta)\tilde{H},
\end{eqnarray}
\end{subequations}
where $<i,j>$ denotes a pair of neighbors that overlap,  $\tilde{z}$ is the number of overlapping neighbors of a $\Gamma_i$, and $\eta$ is the overlapping parameter defined below in Eq.~\eqref{eq:anticomm}, which measures how much the anticommutator can contribute negatively.
We will discuss below in detail how to determine $\eta$.

Importantly, if $1-\tilde{z}\eta>0$, then $\tilde{H}$ is gapped and so is the original AKLT Hamiltonian. In terms of Eq.~(\ref{eq:gammaH}), we can take $\tilde{\gamma}=1-\tilde{z}\eta>0$, and combining with $\gamma_0$, we find a lower bound on the gap of original AKLT Hamiltonian,
\begin{equation}\label{eq:lower}
    \Delta_{\rm lower}=\gamma_0(1-\tilde{z}\eta).
\end{equation}

\subsection{Reducing the gap criterion to an eigenvalue problem}


From the above discussion, we know that $\eta < 1/\tilde{z}$ guarantees the existence of the gap. It is thus important to discuss the property $\eta$,  defined as the greatest possible negative contribution the cross-term of overlapping subgraphs can make,
\begin{equation}
\label{eq:anticomm}
    \{\tilde{H}_i,\tilde{H}_j\}\geq-\eta(\tilde{H}_i+\tilde{H}_j).
\end{equation}
For convenience, we define $E=\mathbb{1}-\tilde{H_i}$, $F=\mathbb{1}-\tilde{H_j}$ as the complements of the subgraph projectors. Equation~\eqref{eq:anticomm} holds if and only if the following holds,
\begin{equation}
\label{eq:anticomm2}
    \{E,F\}\geq-\eta(E+F),
\end{equation}
as proved in Ref.~\cite{Fannes1992} and Ref.~\cite{PomataWei2019}.
Geometrically, $\eta$ is the cosine of the least nontrivial angle between the hyperplanes corresponding to the two projectors $E$ and $F$, and $1\pm\eta$ is the greatest or least non-integer eigenvalue of $E+F$. This can be shown by considering an eigenvector $w$. If we exclude the subspaces $\ker E \bigcap \ker F$ and $\ker E^\bot \bigcap \ker F^\bot$, which correspond to eigenvalues $0$ and $2$, respectively, then there is a unique decomposition of any vector $w$  into two vectors  $w=u+v$, with $u$ and $v$ lying in $\ker F^\bot$ and $\ker E^\bot$, respectively. The two vectors $u$ and $v$ in the decomposition obey~\cite{PomataWei2019} 
\begin{subequations}
    \begin{eqnarray}
    &E u=u, E v=-\alpha u\\
    &F v=v, F u=-\alpha v,
    \end{eqnarray}
\end{subequations}
where $\alpha \in [-1,1]$ is related to an eigenvalue of  $E+F$ and $\{E,F\}$ by
\begin{subequations}
\begin{eqnarray}
    &(E+F)(u+v)=(1-\alpha)(u+v),\\
    &(E F+F E)(u+v)=-\alpha(E+F)(u+v).
\end{eqnarray}
\label{eq:sum-anticomm-eigs}
\end{subequations}
Since $1\pm \alpha$ gives all the non-integer eigenvalues of $\tilde{H}_i+\tilde{H}_j=2\mathbb{1}-E-F$, by comparing \eqref{eq:sum-anticomm-eigs} with the definition of $\eta$ 
 \begin{equation}
    \eta=\sup_{\alpha \notin \mathbb{Z}} |\alpha|,
\end{equation}
as was proven in Ref.~\cite{PomataWei2019}.

\subsection{Projecting the problem into a lower-dimensional subspace}

For a overlapping pair of subgraphs $\Gamma_i$ and $\Gamma_j$, we split their union into three subgraphs $\Gamma_L\equiv\Gamma_i/\Gamma_j$, $\Gamma_M\equiv\Gamma_i\bigcap\Gamma_j$, $\Gamma_R\equiv\Gamma_j/\Gamma_i$. The non-integer eigenvectors of $E+F$ are in the subspace spanned by the states which are the ground states of the local AKLT Hamiltonians for $\Gamma_L$, $\Gamma_M$, and $\Gamma_R$. So we can do the diagonalization on a smaller ``virtual'' space.

To show this, we denote by $A_L$, $A_M$, and $A_R$ as the projectors onto the ground space of AKLT Hamiltonian for $\Gamma_L$, $\Gamma_M$, and $\Gamma_R$, respectively. Without loss of generality we consider $A=A_L$, satisfying~\cite{PomataWei2019,PomataWei2020}
\begin{subequations}
    \begin{eqnarray}
        &&E A=A E=E,\\
        &&F A=A F.
    \end{eqnarray}
\end{subequations}
The first comes from the frustration-freeness of AKLT Hamiltonians and $\Gamma_A\subset \Gamma_i$. The second comes from $\Gamma_A \bigcap \Gamma_j=\emptyset$. Then for $w=u+v$, when the eigenvalue of $E+F$ is noninteger, so $\alpha\neq0$, the projector $A$ preserves $w$,
\begin{equation}
    A w=\alpha^{-2} A F E w=-\alpha^{-2} F A E w=\alpha^{-2} F E w=w.
\end{equation}
Thus $A_L$, $A_M$, and $A_R$ preserve the non-integer spectrum of $E+F$. We can thus find an orthonormal basis which spans the image of $A_L\otimes A_M \otimes A_R$.

\subsection{Constructing the projectors using tensor networks}

 \begin{figure}
 {\centering
 \includegraphics[width=0.45\textwidth]{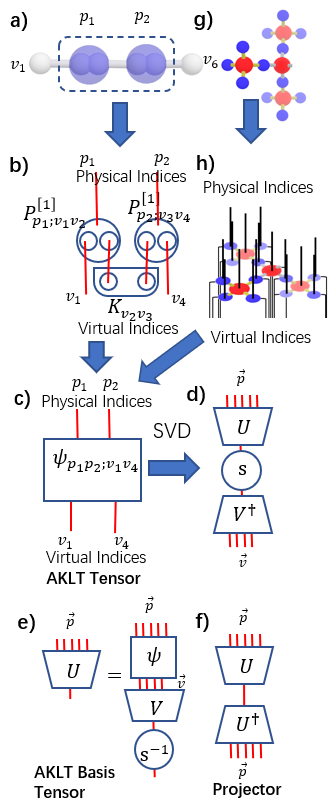}}
 \caption{\label{fig:TensorBuild1}
 Illustration of how to extract the projector to the AKLT ground space of an given region or subgraph.
 (a) A subgraph in a 1D AKLT chain is used as an example, where each physical spin can be written as symmetric sum of virtual spins. 
 (b,c) The AKLT tensor $\Psi=U U^\dagger$ is constructed by acting symmetrizers $P$ at each vertex on the tensor product of antisymmetric virtual spin doublet states $K$ at each edge and external virtual spins at dangling edges.
 (d,e) The AKLT basis tensor $U$ is the orthonormalized AKLT Tensor, using a singular value decomposition $\Psi=U s V^\dagger$. To avoid storing tensors with large physical dimensions explicitly, $s$ and $V$ are calculated numerically from $\Psi^\dagger \Psi$. $U$ is represented in terms of tensor network as $U=\Psi V s^{-1}$.
 (f) The projector $\Pi=U U^\dagger$. 
 (g,h) Another example of an AKLT Tensor on a more complicated subgraph.
} 
 \end{figure}
 
  \begin{figure}
 {\centering
 \includegraphics[width=0.45\textwidth]{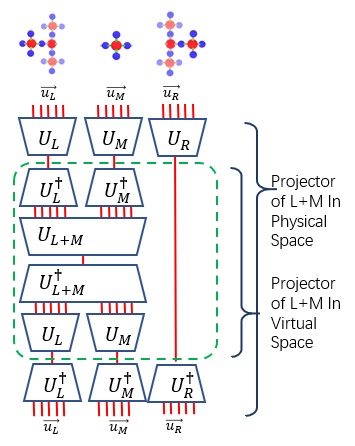}}
 \caption{\label{fig:TensorBuild2}
  We can reduce the dimension of the projector $E=U_{L+M} U_{L+M}^\dagger$ by using the fact that $A_L=U_L U_L^\dagger$,$A_M$,$A_R$ only annihilate eigenvectors of $E$ with integer eigenvalue. Since $U_L$ acts isometrically on the image of $A_L$, the non-integer eigenvalues of $E+F$ can be obtained by performing the spectral decomposition of tensors derived from the tensor in the dashed green box and the corresponding tensor for $M+R$. 
} 
 \end{figure}

There is an isometry that maps the ground space of the AKLT Hamiltonian on a subgraph to the ``virtual space'' which consists of virtual spin degrees of freedom for each vertex with dangling edges~\cite{PomataWei2019,PomataWei2020}. So there is an correspondence from the physical spins on the bulk to the virtual spins on the boundary. Our goal is to replace the physical indices in the projectors $E$ and $F$ by virtual indices, on the boundary of subgraphs $\Gamma_L$, $\Gamma_M$, and $\Gamma_R$. That makes it possible to numerically solve a diagonalization problem with a much larger original physical dimension. We emphasize that the reduction, though done numerically via the singular value decomposition (SVD), is exact in principle. 

This isometry is built out of the AKLT tensor $\Psi_\Gamma$, which can be written in terms of a tensor network. The basic building blocks of the tensor network are:

1. For each vertex $a$ with degree $z_a$, an isometry $P_a^{[z_a/2]}$ which maps the total symmetric space of $z_a$ virtual spin-$1/2$ to the physical spin-$z_a/2$. Here are the expressions for a degree-2 vertex and a degree-4 vertex, respectively,
\begin{subequations}
  \begin{eqnarray}
    P^{[1]}_a&=&\ket{1}\bra{\uparrow\uparrow} 
    +\ket{0}\frac{1}{\sqrt{2}}(\bra{\uparrow\downarrow}+\bra{\downarrow\uparrow}) 
    +\ket{-1}\bra{\downarrow\downarrow}\\
    P^{[2]}_a&=&\ket{2}\bra{\uparrow\uparrow\uparrow\uparrow}\nonumber\\
   &+&\ket{1}\frac{1}{2}(\bra{\uparrow\uparrow\uparrow\downarrow}+\bra{\uparrow\uparrow\downarrow\uparrow}+\bra{\uparrow\downarrow\uparrow\uparrow}+\bra{\downarrow\uparrow\uparrow\uparrow})\nonumber\\
   &+&\ket{0}\frac{1}{\sqrt{6}}(\bra{\uparrow\uparrow\downarrow\downarrow}+\bra{\uparrow\downarrow\uparrow\downarrow}+\bra{\downarrow\uparrow\uparrow\downarrow}\nonumber\\
   &&\;\;\;\;\;\;\;+\bra{\uparrow\downarrow\downarrow\uparrow}+\bra{\downarrow\uparrow\downarrow\uparrow}+\bra{\downarrow\downarrow\uparrow\uparrow})\nonumber\\
   &+&\ket{-1}\frac{1}{2}(\bra{\uparrow\downarrow\downarrow\downarrow}+\bra{\downarrow\uparrow\downarrow\downarrow}+\bra{\downarrow\downarrow\uparrow\downarrow}+\bra{\downarrow\downarrow\downarrow\uparrow})\nonumber\\
   &+&\ket{-2}\bra{\downarrow\downarrow\downarrow\downarrow}.
  \end{eqnarray}
\end{subequations} 

2. For each edge $e=(a,b)$, an antisymmetric tensor $K$ which contracts with a pair of virtual spin-$1/2$'s on each end of the edge,
\begin{equation}
    K_{e}=\frac{1}{\sqrt{2}}(\ket{\uparrow}_a\ket{\downarrow}_b-\ket{\downarrow}_b\ket{\uparrow}_a).
\end{equation}
The above two points, which combined define the AKLT wavefunction, show how it is naturally expressed in terms of a tensor network.

As illustrated in Fig.~\ref{fig:TensorBuild1}(h), if we contract all the vertex tensors and edge tensors, we can treat the resulting tensor as a matrix, with the `right' being a fusion of uncontracted indices that correspond to virtual spins on the dangling edges of the subgraph, and the `left' being a fusion of indices corresponding to the physical spins on the vertices. However, this matrix is not full-rank, because for vertices with  $z_a'>1$ dangling edges, the vertex tensor only acts on the total symmetric subspace of the remaining $z_a'$ spin-$1/2$s. We can `fix' it by further contracting the remaining $z_a'$ virtual spins with $P_a'^{[z_a'/2]\dagger}$, which means that we replace these $z_a'$ spin-$1/2$s with a spin-$z_a'/2$ degree of freedom. It was shown that such a counting is exact~\cite{PomataWei2020} and is related to the uniqueness ground of the AKLT model under appropriate boundary conditions~\cite{KLT}.

Altogether we arrive at the AKLT tensor, as illustrated in Fig.~\ref{fig:TensorBuild1}(a)-(c),
\begin{equation}
    \Psi_\Gamma=\prod_{a\in\Gamma}P_a^{[z_a/2]}\prod_{e\in\Gamma}K_e\prod_{a\in\partial\Gamma}P_a'^{[z_a'/2]\dagger}.
\end{equation}
This tensor $\Psi_\Gamma:\mathcal{H}_{virtual}\rightarrow\mathcal{H}_{physical}$ is an isometry from the virtual space to the AKLT ground subspace of the physical degrees of freedom. Therefore, the ground subspace is the span of the left singular vectors of $\Psi_\Gamma$, which can be obtained from the singular value decomposition (SVD) of $\Psi_\Gamma$,
\begin{subequations}
  
 \begin{eqnarray}
  &&\Psi_\Gamma = U_\Gamma s_\Gamma V_\Gamma^\dagger,\\    &&\Pi_\Gamma=U_\Gamma U_\Gamma^\dagger,
   \end{eqnarray}
\end{subequations}
where $U_\Gamma$ is the orthonormalized AKLT tensor whose column vectors span the ground subspace of the AKLT Hamiltonian $H_\Gamma$;  $\Pi_\Gamma$ is the projector to the ground subspace. 

To avoid the large physical dimension and to take advantage of the smaller virtual dimension, we can perform the SVD or  the eigenvalue decomposition for $\Psi^\dagger \Psi$, which is a $\dim \mathcal{H}_{v}$ by $\dim \mathcal{H}_{v}$ matrix,
\begin{equation}
    \Psi^\dagger \Psi=V_\Gamma s_\Gamma^2 V_\Gamma^\dagger,
    \label{equ:pdp}
\end{equation}
where $U_\Gamma$ is a $\dim \mathcal{H}_{p}$ by $\dim \mathcal{H}_{v}$ matrix, which may be too large to fit in the computer memory. So we express $U_\Gamma$ from the tensor network representation of $\Psi$, as seen in Fig.~\ref{fig:TensorBuild1}(d,e), 
\begin{equation}
    U_\Gamma = \Psi_\Gamma V_\Gamma s_\Gamma^{-1}.
    \label{equ:svd}
\end{equation}

Using the tensor-network representations of $A_L=U_L U_L^\dagger, A_M, A_R, E, F$, we can write the action of $(A_L \otimes A_M \otimes A_R)E$ and $(A_L \otimes A_M \otimes A_R)F$ by contracting the physical indices; see Fig.~\ref{fig:TensorBuild2}. 

To extract the non-integer eigenvalues, we can consider only the tensor inside the dashed box in Fig.~\ref{fig:TensorBuild2}(f). This is equivalent to studying the action of $E$ and $F$ on the image of $A_L\otimes A_M \otimes A_R$, using the orthonormal basis $U_L \otimes U_M \otimes U_R$.

\subsection{Methods to extract the largest non-integer eigenvalue}

 \begin{figure}
 {\centering
 \includegraphics[width=0.45\textwidth]{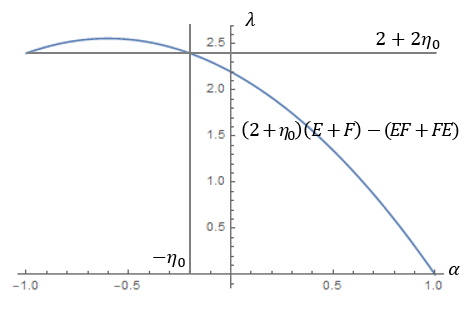}}
 \caption{\label{fig:Nonlinear}
 The $\alpha$ vs. $\lambda$ plane, where $\alpha$ is as in \eqref{eq:sum-anticomm-eigs} and $\lambda$ is the eigenvalue of a quadratic polynomial of $E+F$, which we hope to choose such that eigenvectors with $\alpha\leq-\eta_0$ will yield an eigenvalue $\geq 2+2\eta_0$. Note that $\alpha$ is symmetrically distributed with respect to 0.
 }
 \end{figure}

We use the standard \texttt{ARPACK} library to do the eigenvalue decomposition in Eq.~\eqref{equ:pdp} and sparse maximum-eigenvalue extractions of quadratic polynomials of $E+F$. Below, we mostly focus on the case where the size of $E+F$ (when reduced to virtual degrees of freedom) is large enough that it cannot be exactly diagonalized. 

The tensor-network representations in \eqref{equ:svd} and Fig.~\ref{fig:TensorBuild1},~\ref{fig:TensorBuild2} are handled using a python package called \texttt{tensornetwork}. The action of $E$ on a vector $v\in\mathcal{H}_{v}$ is evaluated by contracting the tensor network representation of this expression. The path of contraction is calculated using the python package \texttt{opt\_einsum}, to ensure the size of intermediate tensors do not exceed the memory limit.

Given the action of $E$ and $F$, one can extract the spectrum using  the iterative method provided in \texttt{ARPACK}. However, the size of the problem makes it only feasible to extract the first several greatest-magnitude eigenvalues. Thus, we consider the quadratic polynomial of $E+F$:
\begin{equation}
    O\equiv(2+\eta_0)(E+F)-(E F+F E).
\end{equation}
The action of $O$ on an eigenvector $w$ is
\begin{equation}
    O w=(2+\eta_0+\alpha)(1-\alpha) w.
\end{equation}
As illustrated in Fig.~\ref{fig:Nonlinear}, our goal is to find $\eta=\sup_{\alpha \notin \mathbb{Z}} |\alpha|$, where the noninteger $\alpha$'s are symmetrically distributed around 0. 

We first guess an $\eta_0$ and evaluate the largest eigenvalue $\lambda$ of $O$. If $\lambda$ is close to $2+2\eta_0$ up to numerical error, then we can almost claim that there is no non-integer $\alpha \gtrsim \eta_0$. In this case, the largest eigenvalue comes from $a=-1$ or $a\approx-\eta_0$. The approximation sign comes from numerical error. We can further exclude the second case by doing another calculation using a slightly smaller $\eta_0'$. If $\lambda'\approx2+2\eta_0'$, then we know $\alpha=-1$. By decreasing $\eta_0$ carefully we finally find an $\eta_0$, where $\lambda-(2+2\eta_0)$ is large enough compared to the numerical error. Then, we have
\begin{equation}
    \alpha=-\frac{1}{2}(1+\eta_0\pm\sqrt{(\eta_0+3)^2-4 \lambda}).
\end{equation}
The $+$ branch is less than $-1$ and can be excluded, so
\begin{equation}
    \eta=-\frac{1}{2}(1+\eta_0-\sqrt{(\eta_0+3)^2-4 \lambda}).
\end{equation}
Importantly, if $1- \tilde{z} \eta >0$, where $\tilde{z}$ is the number of overlapping neighbors of a subgraph,  then the corresponding AKLT model has a nonzero spectral gap.
\subsection{Evaluating the lower bound of the subgraph Hamiltonian gap}

Having obtained the overlapping parameter $\eta$ to verify the gap, the next step is to  give a lower bound on the gap  of the  AKLT Hamiltonian via $\Delta_{\rm lower}=\gamma_0(1-\tilde{z}\eta)$, where  $\gamma_0$ is the   gap  of the weighted AKLT Hamiltonian on the subgraph $\Gamma_i$,
\begin{equation}
    H_{\Gamma_i}=\sum_{e\in\Gamma_i}w_e H_e\geq\gamma_0\tilde{H}_i.
\end{equation}
The weight of the same edge Hamiltonian in different subgraphs sums up to unity: $\sum_i w_e^{(\Gamma_i)}=1$.
We note that the weighted AKLT Hamiltonian on a finite graph is always gaped, since there are only finite number of states. 

There are three ways to calculate the gap of this weighted AKLT Hamiltonian, which we now discuss.

\subsubsection{Direct Method}\label{sec:Direct}

The \texttt{ARPACK} library contains a procedure to get the algebraically smallest eigenvalue given a sparse Hermitian linear operator. To calculate the gap, which is the second smallest eigenvalue above (with the smallest one being zero), we use the AKLT projector to shift the ground states to a higher level, 
\begin{equation}
    OP=\sum_{e\in\Gamma_i}w_e H_e+\big(\sum_{e\in\Gamma_i}w_e\big)\Pi_{\Gamma}.
\end{equation}
Then we can apply the ARPACK package to  calculate the  smallest eigenvalue numerically for the shifted Hamiltonian.
\subsubsection{Lower-bounding with sub-subgraph projectors}
\label{sec:gapVirtual}

For larger subgraphs for which one cannot apply the direct method, we decompose it further into sub-subgraphs which collectively contain all the edges in the subgraph, with weights on each edge summing up to the corresponding weight of the edge in the subgraph; see Fig.~\ref{fig:alternative}(a).

The weighted subgraph AKLT Hamiltonian is the sum of weighted  AKLT terms in the sub-subgraphs, which can be further lower-bounded by the projectors orthogonal to the local ground state with appropriate weights ($w_j'$),
\begin{equation}\label{gap:method2}
    H_{\Gamma_i}=\sum_{\Gamma_j'}H_{\Gamma_j'}\geq\sum_{\Gamma_j'}w_j'\tilde{H}_{\Gamma_j'},
\end{equation}
where $w_j'$ are the gaps of the Hamiltonian on the sub-subgraphs, which can be calculated in method 1, and $\tilde{H}_{\Gamma_j'}=\openone_{\Gamma_j'}-\Pi_{\Gamma_j'}$ is the projector onto the Hilbert space orthogonal to  the local ground state supported in  the sub-subgraph $\Gamma'_j$ of of the subgraph $\Gamma_i$.

Similar to the reduction from the physical to virtual degrees of freedom when we calculate the  $\eta$ parameter, we can project the Hamiltonian on the r.h.s. of Eq.~(\ref{gap:method2}) to the virtual degrees of freedom and calculate its gap in order to bound the gap of the above $H_{\Gamma_i}$;  see Fig.~\ref{fig:alternative}(b).
This method was proposed in Proposition 5 of the SM of Ref.~\cite{PomataWei2020}.

\subsubsection{Overlapping sub-subgraphs}
\label{sec:gapOverlapping}
In Eq.~(\ref{gap:method2}), we divide a subgraph $\Gamma_i$ further into a few overlapping sub-subgraphs $\Gamma_j'$. If these $\Gamma_j'$ are chosen such that they are related by symmetry such as rotation, then the gap of their weighted AKLT Hamiltonians is identical, i.e., $w_j'=w$, yielding 
$H_{\Gamma_i}\ge w\sum_{\Gamma'_j} \tilde{H}_{\Gamma_j'}$. Similar to Eq.~(\ref{eq:gammaH}), we calculate the $\eta$ parameter for $\sum_{\Gamma'_j} \tilde{H}_{\Gamma_j'}$, and a lower bound on the gap of $H_{\Gamma_i}$ can be obtained as $w(1-z\eta)$, where $z$ denotes the number of overlapping neighbors of each sub-subgraph $\Gamma_j'$.
Of course, such a lower bound is valid only when $\eta<1/z$.

\section{Results for three AKLT models}
\label{sec:results}

\subsection{Decorated Diamond Lattice}
\label{diamond}
We choose the subgraph $\Gamma$ as Fig.~\ref{fig:DecoratedDiamondFour}. Each $\Gamma_i$ is overlapping with $\tilde{z}=12$ other $\Gamma_j$s in the same configuration. The virtual dimensions of $\Gamma_i/\Gamma_j$, $\Gamma_i\bigcap\Gamma_j$, $\Gamma_j/\Gamma_i$ are $1024\times16\times1024=16{,}777{,}216$. The parameter $\eta$ is determined from the larger non-integer eigenvalue $1+\eta$ of the sum of the projects $E+F$ is calculated to be $\eta=0.04131015388<\frac{1}{12}$. This shows that the mixed spin-1/spin-2 AKLT model on the singly-decorated diamond lattice has a nonzero gap in the thermodynamic limit.

We note that, in principle, our calculations are accurate to the machine precision. But in the following, we shall only present ten digits of accuracy with the eleventh digit being modified by the round-up of the remaining digits for $\eta$,   and denote the above result as $\eta=0.041310153882$.  However, for the gap estimate, we will modify the 11th digit as the round-down from the remaining digits.

Next, we give two approaches to lower bound the spectral gap of the AKLT Hamiltonian on the singly decorated diamond lattice.
\subsubsection{First approach for the gap lower bound}
\begin{figure}
 {\centering
 \includegraphics[width=0.4\textwidth]{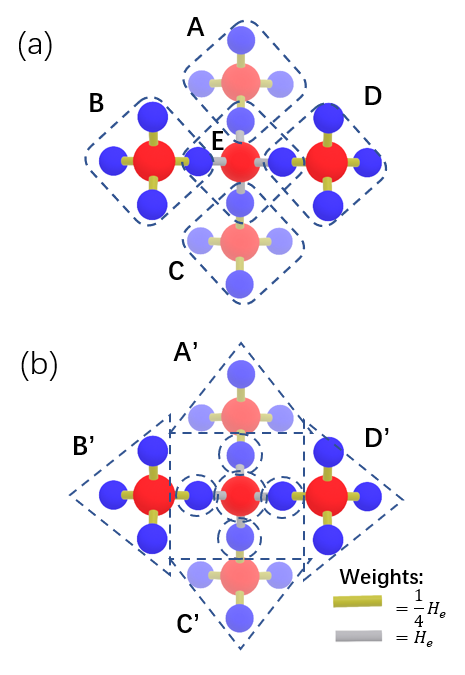}}
 \caption{\label{fig:alternative}
Approach I for lower-bounding the gap. (a) The subgraph is decomposed into four overlapping regions.  (b) An isometry will be applied to the Non-overlapping regions  (four triangular shapes $A'$, $B'$, $C'$ and $D'$) for dimensional reduction. }
\end{figure}

 Here we apply the method described in Sec.~\ref{sec:gapVirtual} to to lower-bound the gap of an AKLT Hamiltonain.  To do this, we consider the lower-bounding Hamiltonian $H_{\rm AKLT}=\sum_i H_{\Gamma_i}\ge \gamma_0 \sum_i \tilde{H}_i$, where $\gamma_0$ is the gap of the weighted AKLT Hamiltonian $H_{\Gamma_i}$ in a region $\Gamma_i$, which consists   of  five overlapping sub-regions $A,B,C,D, E$, as  illustrated in Fig.~\ref{fig:alternative}a.  We note that the weights in front of local AKLT term are 1 in sub-region E and 1/4 in all other sub-regions A-D.  The lower bound of the AKLT gap is $\Delta_{\rm lower}=\gamma_0(1-\tilde{z}\eta)$, where the quantity in the bracket   is already calculated. As we cannot directly calculate $\gamma_0$, due to the excessive Hilbert-space dimension $3^{16}5^5$, we seek a lower bound. To do this, we first lower-bound the Hamiltonian $H_{\Gamma_i}$ via
$H_{\Gamma_i}\ge \gamma_5\Big[ (\openone_E-\Pi_{E})+\sum_{j=A}^D \frac{1}{4}(\openone_j-\Pi_j)\Big]:=\gamma_5 \tilde{H}_5$, where $\Pi_j$ is the projector to the ground space in sub-region $j$, and $\openone_j$ is the identity operator supported on it. The gap of the AKLT Hamiltonian in each region is calculated to be  $\gamma_5=0.17064623273$. 

The second step is to bound the gap of $\tilde{H}_5$. Even though it has the same dimension as that of $H_{\Gamma_i}$, it consists of projectors on sub-regions. Therefore, we can apply an isometry $U_5$, similar to those used in the complexity reduction for the $\eta$ parameter, that consists of a product of isometric transformations on the four non-overlapping sub-regions $A'$, $B'$, $C'$, and $D'$, shown in Fig.~\ref{fig:alternative}b. This reduces the dimension of $\tilde{H}_5$ to that of $\tilde{H}'_5\equiv U_5 \tilde{H}_5 U_5^\dagger$, which is acting on a Hilbert space of dimension  $2^{16}\cdot 3^4\cdot 5$. Such a reduction allows us to use the Lanczos method to show the lowest nonzero eigenvalue of $\tilde{H}_5$ to be $\gamma_R=0.15830084148$.
As this number is smaller than the minimal weight 1/4 of the projectors in $\tilde{H}_5$, $\gamma_R$ is a lower bound on the energy gap of $\tilde{H}_5$, following from Proposition 5 in the Supplemental Material of Ref.~\cite{PomataWei2020}. Thus $\gamma_0$ is lower-bounded via $\gamma_0\ge \gamma_5\gamma_R\approx 0.027013442238$.

So the lower bound on the gap of the AKLT Hamiltonian on the decorated diamond lattice is
\begin{eqnarray}
    \Delta_{\rm lower}&\geq&{\gamma}_0(1-\tilde{z}\eta)\nonumber\\
    &=&0.013622288769.
\end{eqnarray}

\subsubsection{A second approach for the lower bound}
 Here we give an alternative to obtain the gap lower bound, described in Sec.~\ref{sec:gapOverlapping}. We first give the bound obtained:
\begin{eqnarray}
     && \Delta_{\rm lower}\geq \gamma_0(1-\tilde{z} \eta)\nonumber\\
       &&\geq
    0.013110607533\times(1-12\times 0.041310153882)\nonumber\\
 &&=0.0066113929572.
\end{eqnarray}
 \begin{figure}
 {\centering
 \includegraphics[width=0.45\textwidth]{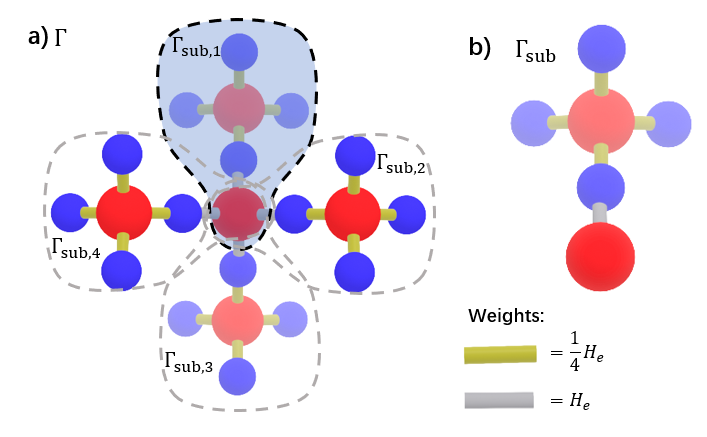}}
 \caption{\label{fig:sublattice}
 Approach II for the gap lower bound via decomposing one subgraph $\Gamma$ into four overlapping sub-subgraphs. }
 \end{figure}
As we explain above,
the gap $\gamma_0$  of the AKLT Hamiltonian on subgraph $\Gamma$ cannot be directly calculated.   The second approach to the decorated diamond case is to    further decomposition of  a subgraph $\Gamma$ into four overlapping sub-subgraphs $\Gamma_{sub,j}$; see Fig.~\ref{fig:sublattice}(a). These four sub-subgraphs are related to each other by rotating with respect to the center spin-2 site.
We thus can lower bound the Hamiltonian
$H_{\Gamma}\ge \gamma_1 \sum_{j=1}^4 \tilde{H}_{sub,j}:=\gamma_1 \tilde{H}_{sub}$, where $\gamma_1$ is the gap of the weighted AKLT Hamiltonian in each region and $\tilde{H}_{sub,j}$ is the projector onto the Hilbert space orthogonal to the local ground space.
The physical dimension of each $\Gamma_{sub,j }$ is $5^2 3^4=2025$, so we obtain the sub-subgraph gap
 $\gamma_1=0.044374363959$ by exact diagonalization.

To obtain a lower bound on the gap of $\tilde{H}_{sub}$, we consider its square,
\begin{subequations}
    \begin{eqnarray}
        \tilde{H}_{sub}^2&\geq&\tilde{H}_{sub,1}+\tilde{H}_{sub,2}+\tilde{H}_{sub,3}+\tilde{H}_{sub,4}\nonumber\\
        &&+\{\tilde{H}_{sub,1},\tilde{H}_{sub,2}\}+\{\tilde{H}_{sub,1},\tilde{H}_{sub,3}\}\nonumber\\
        &&+\{\tilde{H}_{sub,1},\tilde{H}_{sub,4}\}+\{\tilde{H}_{sub,2},\tilde{H}_{sub,3}\}\nonumber\\
       &&+\{\tilde{H}_{sub,2},\tilde{H}_{sub,4}\}+\{\tilde{H}_{sub,3},\tilde{H}_{sub,4}\}\\
    &\geq&(1-z_1 \eta_1) \tilde{H}_{sub},
   \end{eqnarray}
\end{subequations}
where the parameter $\eta_1$ given by
\begin{equation}
\eta_1=\sup_{\alpha\notin\mathbb{Z}}|\alpha|,
\end{equation}
 where $(1-\alpha)$'s are the eigenvalues of the following equation
 \begin{equation}(\tilde{H}_{sub,1}+\tilde{H}_{sub,2})w=(1-\alpha)w.
\end{equation}

 Each $\Gamma_{sub,i}$ overlaps with $z_1=3$ other $\Gamma_{sub,j}$'s, with virtual dimensions $16*5*16=1280$. The parameter
 $\eta_1=0.23484848485$ is calculated using the method described above in Sec.~\ref{sec:method}.
Thus, we obtain a lower bound  $\gamma_0$ on the gap of the subgraph Hamiltonian,
\begin{equation}
    \begin{split}
       & \gamma_0\geq \gamma_1(1-z_1 \eta_1)\\
    &=
    0.044374363959\times(1-3\times 0.23484848485)\\
    &=0.013110607533.
    \end{split}
\end{equation}

We also used three other approaches to lower bound the gap of the singly decorated diamond AKLT Hamiltonian and we list the results from all five different approaches (based on three different partitions of the Hamiltonian) in Appendix~\ref{sec:fivemethods}.

\subsection{Triangle-Octagon Lattice}

We use the subgraph $\Gamma$, as shown in Fig.~\ref{fig:OctagonTriangleLattice}. Each $\Gamma_i$ is overlapping with $\tilde{z}=4$ other $\Gamma_j$'s. The virtual dimension of $\Gamma_i/\Gamma_j$, $\Gamma_i\bigcap\Gamma_j$, $\Gamma_j/\Gamma_i$ are $512\times16\times512=4{,}194{,}304$. This configuration yields an 
$\eta=0.22524594477<\frac{1}{4}$.

The physical dimension of a single subgraph is $5^9=1{,}953{,}125$, which is small enough for a direct spectrum decomposition in the physical space. In practice, we adopt a shifted Hamiltonian $H_\text{shifted}=\sum_{e\in\Gamma}w_e H_e+(\sum_{e\in\Gamma}w_e)\Pi_\Gamma$, which shifts the ground state to an eigenvalue $\sum_{e\in\Gamma}w_e=10$ much larger than the possible gap. As described in Sec.~\ref{sec:Direct}, we then calculate the gap by extracting the least eigenvalue of $H_\text{shifted}$ using \texttt{ARPACK}, which gives
$\gamma_0=0.09764599552$.
The lower bound of the AKLT Hamiltonian on the triangle-octagon lattice is thus
\begin{eqnarray}
 \Delta_{\rm lower}&=&\gamma_0(1-\tilde{z}\eta)\nonumber\\
    &=&
    0.0096685374671.
\end{eqnarray}

\begin{figure}
 {\centering
 \includegraphics[width=0.5\textwidth]{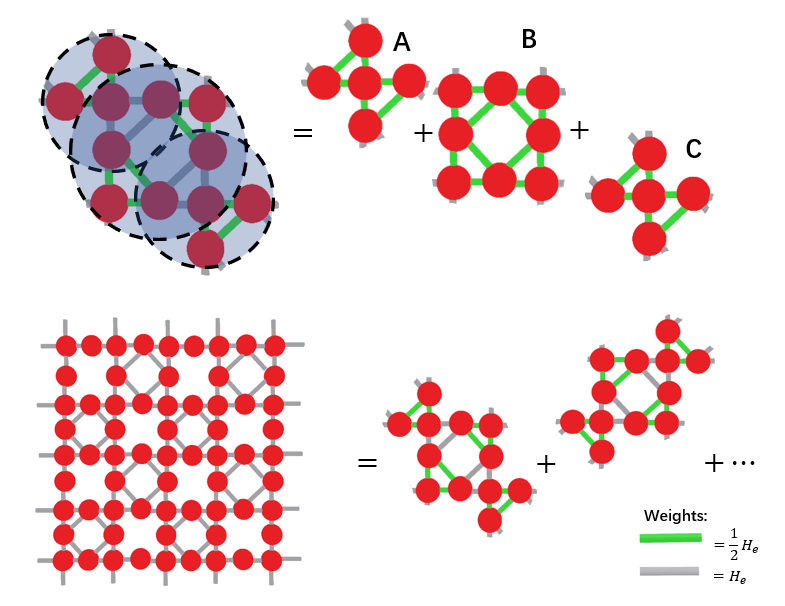}}
 \caption{\label{fig:WeightSum}
The decomposition scheme of a subgraph of the inscribed square lattice. It also shows how the weights of edges of sub-subgraphs sum up to the weights in the subgraph, and the weights of the subgraphs sum up to 1 in the whole lattice.
}
\end{figure}
\subsection{Inscribed Square Lattice}

We use the subgraph $\Gamma$ in Fig.~\ref{fig:InscribedSquareLattice}. Each $\Gamma_i$ overlaps with $\tilde{z}=4$ other $\Gamma_j$'s with the same configuration. The virtual dimension of $\Gamma_i/\Gamma_j$, $\Gamma_i\bigcap\Gamma_j$, $\Gamma_j/\Gamma_i$ are $3888\times 27\times 324=34{,}012{,}224$. The result is
$\eta=0.20517748800<\frac{1}{4}$.

We further decompose the subgraph according to Fig.~\ref{fig:WeightSum}. The gaps of sub-subgraphs are 
$\gamma_A=\gamma_C=0.077207219973$ and
$\gamma_B=0.082508095136$. The physical dimension of the subgraph is $5^{12}=244{,}140{,}625$, which is reduced further (by projection to the virtual degrees of freedom) to $16\times27\times25\times27\times16=4{,}665{,}600$. Using the method described in Sec.~\ref{sec:gapVirtual}, the lower bound of the subgraph gap is calculated to be 
${\gamma}_0=0.058117906479$.

Combining the above results, we have that the lower bound on the gap of the AKLT Hamiltonian on the inscribed square lattice is
\begin{eqnarray}
\Delta_\text{\rm lower}&=&{\gamma}_0(1-\tilde{z}\eta)\nonumber\\
    &=&
    0.010419962243.
\end{eqnarray}

\section{Considerations for the kagome and square lattices}\label{KagomeSquare}

\subsection{Attempts on the Kagome Lattice}

 \begin{figure}
 {\centering
 \includegraphics[width=0.35\textwidth]{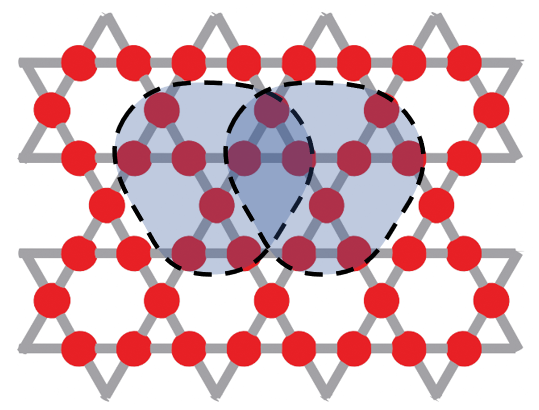}}
 \caption{\label{fig:KagomeLattice}
 The kagome lattice and a overlapping scheme that has been tested.}
 \end{figure}

 We choose the subgraph $\Gamma$ as in Fig.~\ref{fig:KagomeLattice}. Each $\Gamma_i$ overlaps with $\tilde{z}=6$ other $\Gamma_j$'s with the same configuration. The virtual dimension of $\Gamma_i/\Gamma_j$, $\Gamma_i\bigcap\Gamma_j$, $\Gamma_j/\Gamma_i$ are $324\times27\times324=2{,}834{,}352$. The resulting overlap parameter is
 $\eta=0.17067852083>\frac{1}{6}$, which does not satisfy the gap criterion. So with this overlapping scheme we cannot prove the existence of a gap in the kagome AKLT model, although it does not imply the gap does not exist.
 
 Noting that $\eta$ just slightly exceeds the threshold, we naturally guess that by using a larger subgraph partition, we might be able to find an $\eta$ which satisfied the criteria.

 \begin{figure}
 {\centering
 \includegraphics[width=0.35\textwidth]{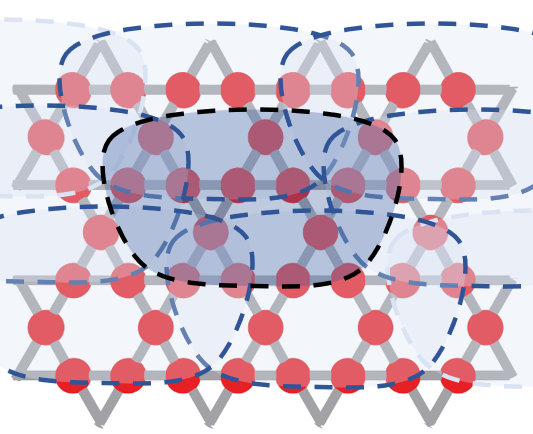}}
 \caption{\label{fig:KagomeLattice2}
 Another kagome lattice overlapping scheme.}
 \end{figure}

 Here we propose another overlapping scheme shown in Fig.~\ref{fig:KagomeLattice2}. For each subgraph there are 6 others that overlap it, and these overlapping pairs are divided into 2 types, noting that some of the pairs are topologically identical. The virtual dimensions are $8748\times27\times8748=2{,}066{,}242{,}608$, $8748\times729\times3888=24{,}794{,}911{,}296$, respectively. Unfortunately, these dimensions are too large for current computing resources.
 
 \begin{figure}
 {\centering
 \includegraphics[width=0.3\textwidth]{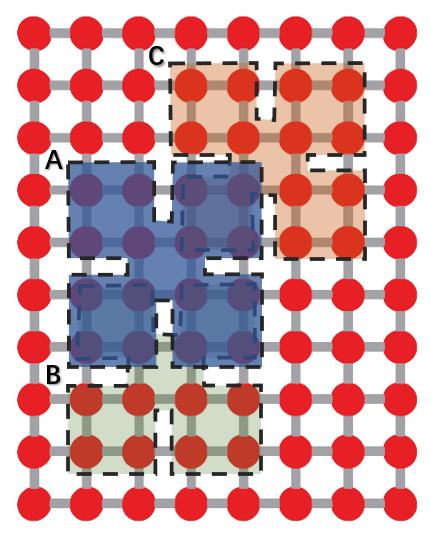}}
 \caption{\label{fig:SquareGuess}
A proposed overlapping scheme for the square lattice. There are two types of overlapping pairs: AB and AC. We suspect the partitioning presented here could be used to prove the existence of the gap of the spin-2 square-lattice AKLT model. However, its computational cost is still out of reach given present resources. 
}
\end{figure}

\subsection{A possible attempt on the square lattice}

Now we present a potentially useful overlapping scheme for the square lattice; see Fig.~\ref{fig:SquareGuess}. Each subgraph overlaps with $\tilde{z} =8$ neighboring subgraphs. The overlapping pairs can be divided into 2 types: AB (4 pairs) and AC (4 pairs). The virtual dimension of AB is $2916\times6561\times2916=55{,}788{,}550{,}416$. Of AC it is $78732\times81\times78732=502{,}096{,}953{,}744$. We hope that it could be proven that $4\eta_{AB}+4\eta_{AC}<1$, which would prove the existence of a spectral gap in the square lattice AKLT model.

\section{A finite-size criterion for the square-lattice model}
\label{Finite}
\begin{figure}
 {\centering
 \includegraphics[width=0.4\textwidth]{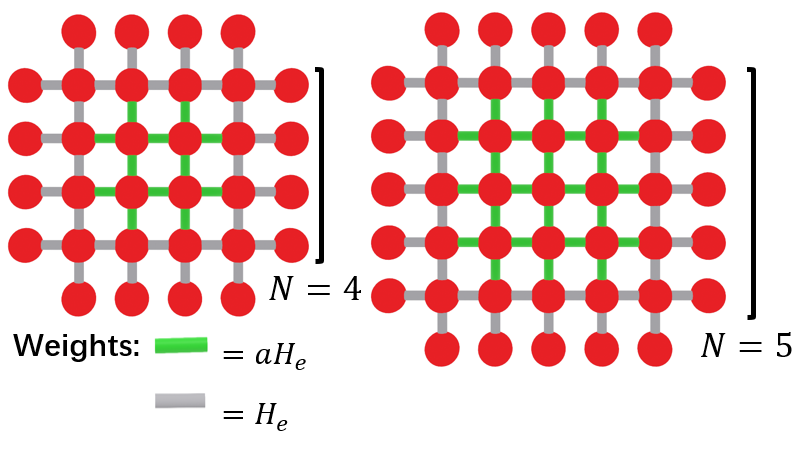}}
 \caption{\label{fig:square4x4}
An illustration of the subgraph $\mathcal{F}_\square$, with $N\times N$ squares and $4N$ surrounding sites. 
We note that  $N\geq 3$, and two examples with $N=4$ and $N=5$ are shown.}
\end{figure}

Here we prove a finite-size criterion, inspired by the work of Lemm, Sandvik and Wang~\cite{LemmSandvikWang2020} on the hexagonal lattice. To do this, we select an $N\times N$ region with $4N$ additional sites around it, as shown in Fig.~\ref{fig:square4x4}, and a factor $a$ to weight Hamiltonian terms by.  We find that the original AKLT gap in the thermodynamic limit can be bounded by the following expression:
\begin{equation}
\label{eq:LSWgap}
    \Delta\geq\frac{f(a)}{g(a)}\left(\gamma_F(a)-\frac{f(a^2)-g(a)}{f(a)}\right),
\end{equation}
where
\begin{eqnarray}
     f(a)&\equiv&2(2N-1)+(N-1)(N-2) a,\\
     g(a)&\equiv&2N+2(N-2)a+(N-2)^2 a^2,
\end{eqnarray}
and $\gamma_F(a)$ is the actual gap of the finite-size weighted AKLT Hamiltonian on the subgraph. By fine-tuning the parameter $a$ we may find that the finite-size gap  $\gamma_F(a)$ exceeds the threshold $\gamma_{\rm TH}(a)\equiv\frac{f(a^2)-g(a)}{f(a)}$, making the lower bound a positive value. If so, this   proving  the existence of  the spectral gap for the original AKLT model on the square lattice.
The minimum $\gamma_{\rm TH}$'s for subgraphs with different sizes $N$ (i.e. the feature length) are shown in Table~\ref{tableI}. 
\begin{table}[h]
    \centering
    \begin{tabular}{|c|c|c|}
        \hline
         $N$ & $a$ & $\gamma_{\rm TH}$ \\
         \hline
         4 & 1.28759 & 0.191729 \\
         5 & 1.31366 & 0.156829 \\
         10 & 1.3654 & 0.081199 \\
         20 & 1.39034 & 0.0410889 \\
         100 & 1.40954 & 0.00827357 \\
         \hline
    \end{tabular}
    \caption{\label{tableI}Threshold of the subgraph gap lower bound $\gamma_{\rm TH}$ in order to establish the gap in the thermodynamic limit for the square-lattice AKLT model.}
    \label{tab:gapThresholds}
\end{table}

In fact, by examining its dependence on $N$, we  observe that $\gamma_{\rm TH}\sim O(\frac{1}{N})$. We expect that, as $N$ increases, $\gamma_F$ will converge to $a\Delta$, and so with $N$ large enough it should exceed the threshold, hopefully while the problem size is numerically accessible.
Below, we give the essential details of the proof.

\subsection{ Details for the finite-size criterion }

\def \Hfs{H_{\mathcal{F}_\square}}
\def \HTfs{\tilde{H}_{\mathcal{F}_\square}}
\def \Qfs{Q_{\mathcal{F}_\square}}
\def \Rfs{R_{\mathcal{F}_\square}}
\def \Efs{\mathcal{E}_{\mathcal{F}_\square}}

We use $\mathcal{F}_\square$  to denote an instance of the weighted graph $\mathcal{F}$ as a subgraph of the lattice  $\Lambda$ in Fig.~\ref{fig:square4x4}, which consists of $N\times N$ plaquettes at the center, including a central plaquette $\square\in\Lambda$, and $4N$ surrounding `dangling' sites connecting to it.  The edge set of $\mathcal{F}_\square$ is denoted by $\Efs$.

To each translation of $\mathcal{F}_\square$, indexed by all plaquettes $\square$, we assign an operator
$\Hfs=\sum_{e\in\mathcal{E}_{\mathcal{F}_\square}} w_e P_e$, where $P_e$ is the AKLT Hamiltonian term (a projector) on two neighboring spins connected by an edge $e$ and the weight $w_e$ is either $1$ or $a$, according to the pattern indicated in Fig.~\ref{fig:square4x4}. We then square $\Hfs$  and sum over all translations (of plaquette $\square$),
\begin{equation}
    \mathcal{A}\equiv\sum_{\square\in\Lambda}\Hfs^2.
\end{equation}
There are two operator inequalities that we will derive, following the idea in Ref.~\cite{LemmSandvikWang2020}:
\begin{eqnarray}\label{equ:OtherAGeq}
   && \mathcal{A}\geq f(a)\gamma_{\mathcal{F}}H,\\
\label{equ:OtherALeq}
   && \mathcal{A}\leq f(a^2) H+g(a)(Q+R),
\end{eqnarray}
where $f$ and $g$ are two functions defined below, $H$ is the original AKLT Hamiltonian on the whole lattice $\Lambda$, and $Q$ and $R$ contain terms involving  pairs of edges in $H^2$  which share a vertex or not, respectively:
\begin{align}
    H&=\sum_{e\in\mathcal{E}}P_e,\\
    Q&=\sum_{e,e'\in\mathcal{E}, e\sim e'}\{P_e,P_{e'}\},\\
    R&=\sum_{e,e'\in\mathcal{E}, e\cancel{\sim}e'}\{P_e,P_{e'}\}.
\end{align}
After we square the total Hamiltonian $H$, the squared terms give back $H$ and there are two types of cross-terms, such that
\begin{equation}
    H^2=H+Q+R.
\end{equation}
By combining Eqs.~\eqref{equ:OtherAGeq} and~\eqref{equ:OtherALeq}, derived below, we conclude that
\begin{equation}
\label{eq:boundH2}
    H^2\ge\frac{f(a)}{g(a)}\left(\gamma_F(a)-\frac{f(a^2)-g(a)}{f(a)}\right)H,
\end{equation}
and hence the lower bound in Eq.~(\ref{eq:LSWgap}), provided the expression inside the bracket in Eq.~(\ref{eq:boundH2}) is positive.

{\it Proof of Eqs.~\eqref{equ:OtherAGeq} and~\eqref{equ:OtherALeq}.}
We first study the number of single-edge terms in $\sum_{\square\in\Lambda}\Hfs$. There are two equivalent types of edges, vertical horizontal edges, each of which appears, in $\mathcal{F}_\square$, $(N-1)(N-2)$ times within the central square and $2(2N-1)$ times outside of it. Thus, by translation, the accumulated weight for each edge term is $f(a)=2(2N-1)+(N-1)(N-2) a$. If we label the gap of $\Hfs$  as $\gamma_\mathcal{F}$, then we can lower-bound $\mathcal{A}$ as
\begin{equation}
    \mathcal{A}=\sum_{\square\in\Lambda}\Hfs^2\geq\sum_{\square\in\Lambda}\gamma_\mathcal{F}\Hfs=\gamma_\mathcal{F}f(a)H.
\end{equation}

We then consider the number of cross-terms in $\mathcal{A}=\sum_{\square\in\Lambda}\Hfs^2$ and decompose $\Hfs^2=\HTfs+\Qfs+\Rfs$, where
\begin{subequations}
\begin{eqnarray}
\label{equ:finiteSizeHQR}
   \HTfs&=&\sum_{e\in\Efs}w_e^2P_e,\\
  \Qfs&=&\sum_{e,e'\in\Efs, e\sim e'}w_e w_{e'}\{P_e,P_{e'}\},\\
     \Rfs&=&\sum_{e,e'\in\Efs, e\cancel \sim e'}w_e w_{e'}\{P_e,P_{e'}\}.
     \end{eqnarray}
\end{subequations}
Since each cross-term only arises when both edges are in the same subgraph $\mathcal{F}_\square$, we expect we can use constant coefficients to bound $\Qfs$ and $\Rfs$ relative to $Q$ and $R$.

As the weight of each edge in $\HTfs$ is squared,  we straightforwardly determine that the coefficient (which we also call the ``accumulated weight'') for each edge in $\mathcal{A}$ is $f(a^2)$:  
\begin{equation}
    \sum_{\square\in\Lambda}\HTfs=f(a^2)H.
\end{equation}

For each pair of edges which share one vertex, one can easily see by counting that both the parallel and perpendicular cases have the same accumulated weight $g(a)\equiv 2N+2(N-2)a+(N-2)^2 a^2$, and thus
\begin{equation}
    \sum_{\square\in\Lambda}\Qfs=g(a)Q.
\end{equation}

\begin{figure}
 {\centering
 \includegraphics[width=0.4\textwidth]{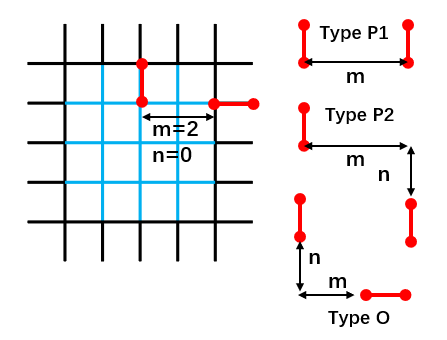}}
 \caption{\label{fig:pairTypes}
 Illustration of the subgraph $\mathcal{F}_\square$ (left) and  three types of edge pairs (right). To analyze the $R$ term in Eq.~(\ref{equ:finiteSizeHQR}), we categorize pairs of edges into three types. The accumulated weight of each type of pair is summarized in Tables~\ref{tab:typeP1},~\ref{tab:typeP2} and~\ref{tab:typeO}, respectively.
 An example pair of type {\bf O}, with $m=2, n=0$ and weight $a\times1=a$, is shown in a $N=5$ subgraph.}
\end{figure}

It turns out that the number of combinations of edges in each class of $R$ can be bounded by those of $Q$. There are three types of equivalent classes of pairs:
\begin{itemize}
    \item[\bf P1:]  a pair of parallel edges separated by $m$ edges along the direction perpendicular to them.
    \item[\bf P2:] a pair of parallel edges separated by $m$ edges in the parallel and $n$ in the perpendicular direction. 
    \item[\bf O:] a pair of orthogonal edges separated by $m$ edges parallel to the first one and $n$ edges parallel to the second.
\end{itemize} See Fig.~\ref{fig:pairTypes} for an illustration of these three classes. We tabulate all possible cases $i$ of the accumulated weight $h_i(a)$ in these three classes in  Tables~\ref{tab:typeP1},~\ref{tab:typeP2} and~\ref{tab:typeO}. As can be checked, for $N\geq 4$ all cases are less than or equal to $g(a)$ for positive $a$, so we conclude that
\begin{equation}
    \sum_{\square\in\Lambda}\Rfs\leq g(a)R.
\end{equation}
Summing up, we arrive at Eq.~\eqref{equ:OtherALeq}.

\begin{table*}[]
    \centering
    \begin{tabular}{|c|c|}
    \hline
        $m$  & $h_i(a)$ \\
    \hline
        $1\leq m \leq N-2$ & $2(N-m)+2(N-1)a+(N-1)(N-2-m)a^2$   \\
        $N-1$ & $N+1$ \\
    \hline
    \end{tabular}
    \caption{The accumulated weights of type {\bf P1} edge pairs in an $N\times N$ subgraph. Note that the case of $m=0$ is excluded, as it counts   terms in $\HTfs$ and gives $f(a)$.  It is straightforward to check that all entries on the right column are not greater than $g(a)=2N+2(N-2)a+(N-2)^2a^2$.}
    \label{tab:typeP1}
\end{table*}

\begin{table*}[]
    \centering
    \begin{tabular}{|c|c|c|}
    \hline
        $m$ & $n$ & $h_i(a)$ \\
    \hline
        $0$ & $1\leq n \leq N-2$  & $2(N-n)+2(N-2)a+(N-2)(N-2-n)a^2$ \\
        $1\leq m \leq N-2$ & $0\leq n \leq N-2$ & $2+2(2N-3-m-n)a+(N-2-m)(N-2-n)a^2$\\
        $N-1$ & $0\leq n \leq N-2$ & $N-n$\\
        $0$ & $N-1$ &$N$\\
        $1\leq m \leq N-2$ & $N-1$ & $N-m$\\
        $N-1$ & $N-1$ & $1$\\

    \hline
    \end{tabular}
    \caption{The accumulated weights of type {\bf P2} edge pairs in an $N\times N$ subgraph. The case $m=n=0$ is excluded as it counts terms in $\HTfs$ and gives $f(a)$. 
    It is straightforward to check that all entries on the last column are not greater than $g(a)=2N+2(N-2)a+(N-2)^2a^2$.}
    \label{tab:typeP2}
\end{table*}

\begin{table*}[]
    \centering
    \begin{tabular}{|c|c|c|}
    \hline
        $m$ & $n$ & $h_i(a)$ \\
    \hline
        $0$ & $1\leq n \leq N-2$ & $(N+1-n)+(3N-5-n)a +(N-2)(N-2-n)a^2$\\
        $0$ & $N-1$ & $N$\\
        $1\leq m \leq N-2$ & $1\leq n \leq N-2$ & $2+2(2N-3-m-n)a+(N-2-n)(N-2-m)a^2$\\
        $1\leq m \leq N-2$ & $N-1$ & $N-m$\\
        $N-1$ & $1\le n \le N-2$ &  $N-n$\\
        $N-1$ & $N-1$ & $1$\\

    \hline
    \end{tabular}
    \caption{The accumulated weights of type {\bf O} edge pairs in an $N\times N$ subgraph. Note that $m$ and $n$ are equivalent, so the table is symmetric under exchange of $m$ and $n$. The case
    $m=n=0$ is excluded as it counts the terms in $\Qfs$ and will give $g(a)$. 
    It is straightforward to check that all entries on the right column are not greater than $g(a)=2N+2(N-2)a+(N-2)^2a^2$.
    \label{tab:typeO}}
\end{table*}

\subsection{Discussion}

For a subgraph with size $N$ and interior edge weight $a$, we can prove that the infinite lattice is gapped if the gap is greater than $\gamma_{TH}(a)$, where
\begin{equation}
    \gamma_{TH}(a)=\frac{(N-2)a^2-2(N-2)a+2(N-1)}{(N-1)(N-2)a+2(2N-1)}.
\end{equation}

\begin{figure}
 {\centering
 \includegraphics[width=0.48\textwidth]{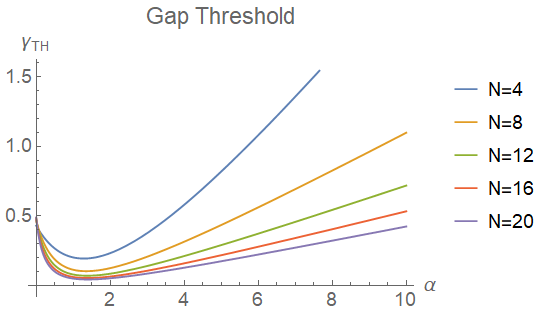}}
 \caption{\label{fig:FiniteSquareGapThreshold}
    The gap threshold $\gamma_{TH}(a)$ for a few finite sizes $N$.}
\end{figure}
The behavior of $\gamma_{TH}(a)$ is shown in Fig.~\ref{fig:FiniteSquareGapThreshold} for a few $N$'s and the minimum values are tabulated in Table~\ref{tab:gapThresholds}.

One can analytically find the minimal value of $\gamma_{TH}$:
\begin{subequations}
    \begin{align}
       a_0\equiv&{\arg\min}_a(\gamma_{TH}(a))\notag\\
       =&\frac{-4N+2+\sqrt{2N^4-2N^3+6N^2-2N}}{(N-1)(N-2)}\\
       \gamma_{TH}&(a_0)\notag\\
       =&-2\frac{N^2+N-\sqrt{2N^4-2N^3+6N^2-2N}}{(N-1)^2(N-2)}.
   \end{align}
\end{subequations}
In the large $N$ limit, $\gamma_{TH}$ is inversely proportional to $N$,
\begin{subequations}
\begin{eqnarray}
    a_0&=&\sqrt2+\mathcal{O}(\frac1N)\\
        \gamma_{TH}(a_0)&=&\frac{2\sqrt2-2}{N}+\mathcal{O}(\frac{1}{N^2}).
    \end{eqnarray}
\end{subequations}
In the large $a$ limit, $\gamma_{TH}$ is linear in $a$ as expected and inversely proportional to $N$:
\begin{equation}
    \gamma_{TH}=\frac{1}{N-1}a+\mathcal{O}(1).
\end{equation}

If the infinite lattice has a gap $\Delta$, one would expect, as one increases $N$, the boundary effect diminishes and the subgraph gap $\gamma_F$ converges to $a\Delta$. However, one can make the gap threshold $\gamma_{TH}\sim\mathcal{O}(\frac1N)$ arbitrary small. Therefore, with a large enough $N$, one should be able to find a configuration where the gap is greater than the threshold, which would prove the existence of the gap.

In contrast, if the infinite lattice is gapless, then one would expect the gap $\gamma_F$ of a finite-size subgraph converges to zero as $N$ increases, and $\gamma_{TH}$ would provide an upper bound for the diminishing gap. However, it is strongly believed that the AKLT model on the square lattice is gapped. In particular, two estimates of the gap in the thermodynamic limit using numerical  tensor-network methods give a consistent value $\Delta\approx 0.06$~\cite{Numerical1,Numerical2}.

\section{Concluding remarks}\label{sec:conclusion}
 We have established the spectral gap for AKLT models on three lattices: (1) the singly decorated diamond lattice; (2) the triangle-octagon lattice, where an octagon is inserted into each plaquette, creating four triangles surrounding each site of the original square lattice; (3) the `inscribed' square lattice, where a diamond (or alternatively circle) is inscribed in every other plaquette of the square lattice.
 
 The first model is composed  of a mixture of spin-2 and spin-1 degrees of freedom.  The spin-2 model on the undecorated diamond lattice is known to be magnetically disordered, but the existence of a gap is still open. The consideration of the decorated diamond lattice may be regarded as an effort towards that as well as a nontrivial three-dimensional model by itself. Intuitively, decorating the diamond lattice by a spin-1 entity on every edge introduces more quantum fluctuation (than the original diamond lattice) and reduces the tendency towards magnetic ordering. The unique ground state and the proof of a gap for the decorated diamond lattice support this intuition. In addition to its existence, we also provide different approaches to lower-bound the value of the gap, though we believe that values we obtained are much smaller than the actual gap.

The other two planar models we considered derive from modification of the square lattice and both consist of uniformly spin-2 entities. To our knowledge, these are the only such AKLT models where a gap has been proven. In the square and kagome lattice models, the gap is believed to exist but  has not been proven. 

We have also made an attempt on the kagome case, and have selected the lattice partition in Fig.~\ref{fig:KagomeLattice}, where $\tilde{z}=6$.  The $\eta$  parameter for such a configuration was calculated to be $\eta\approx 0.1707 > 1/6$. This value unfortunately barely exceeds $1/\tilde{z}$ by less than 3\%. 
One thus needs to consider a partitioning with larger unit cells, such as one shown in Fig.~\ref{fig:KagomeLattice2}. However, the problem size for that is beyond our numerical capability.

For the square lattice, we also suggest the partitioning as shown in Fig.~\ref{fig:SquareGuess} might be used to demonstrate the nonzero gap for the AKLT model on the square lattice. However, the computer memory needed to perform the calculation is also beyond our capability. 

As another approach, one may consider deriving a finite-size criterion like the one used by Lemm, Sandvik and Wang~\cite{LemmSandvikWang2020} in the honeycomb case, and extend their approach to the square lattice.  We have done this  and derived a corresponding criterion for the square lattice. Interestingly, the threshold that the finite-size gap must exceed in order to establish the nonzero spectral gap scales inversely proportional to the linear size of the finite region. If the square-lattice AKLT model possesses a nonzero thermodynamic gap, then as long as one could employ the computational resources to investigate the gap for a sufficiently large finite-size problem, the gap problem could be solved. 

\medskip \noindent {\bf Acknowledgments.} This work was supported by the
National Science Foundation under Grant No. PHY 1915165.

\appendix
 \section{Different lower-bounding methods for the singly decorated diamond lattice}
 \label{sec:fivemethods}
 
 \begin{figure}
 {\centering
 \includegraphics[width=0.3\textwidth]{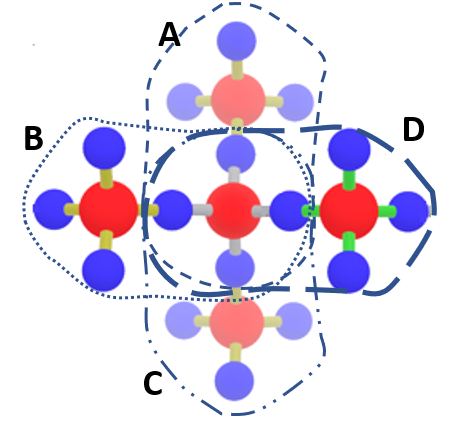}}
 \caption{\label{fig:9spins}
 Partitioning the subgraph into four overlapping regions, each with 9 spins.}
 \end{figure}
 
 In calculating the gap bound for the decorated diamond lattice we  have 5 results coming from 3 different intermediate Hamiltonians. The results here are presented with fewer digits of precision than in the main text.

\renewcommand{\labelenumi}{\Roman{enumi}}

\begin{enumerate}
\item    Five-vertex Hamiltonian terms (a spin-2 plus the four adjacent spin-1s), as in Fig.~\ref{fig:alternative}: $H^5 = \frac{1}{4}(H_A^5+H_B^5+H_C^5+H_D^5)+H_E^5$. The $H_X^5$ are bounded relative to the original Hamiltonian by $\gamma_5=0.1706462$, and the bound of $H_5$ relative to the full 17-vertex projector is $\gamma_R=0.1583008$, giving an overall relative bound of $\gamma_0=0.02701344$.
 
\item  Six-vertex Hamiltonian terms (an outer spin-2 and the four surrounding spin-1s plus the inner spin-2), as in Fig.~\ref{fig:sublattice}: $H^6=H^6_A+H^6_B+H^6_C+H^6_D$, where we have to bound $H^6_X$ relative to sum of terms from the original Hamiltonian where (unlike in the other cases where all coefficients are 1) the four terms that include the outer spin-2 have coefficient 1/4; this gives $\gamma_6'=0.04437436$.
\begin{enumerate}
        \item By computing the overlap parameter $\eta'=0.2348484\ldots$ (between projectors of the two overlapping regions) we get the relative bound $\gamma_0=0.01311061$.
        \item Alternatively, by using Prop 5 of the SM of Ref.~\cite{PomataWei2020} we bound $H^6$ relative to the full projector with $\gamma_R=0.3274050$, getting $\gamma_0=0.01452839$.
\end{enumerate}
        
\item   Nine-vertex Hamiltonian terms (an outer spin-2 and the central spin-2 plus the 7 spin-1s neighboring either), as in Fig.~\ref{fig:9spins}: $H^9 = \frac{1}{4}(H_{A}^9+H_{B}^9+H_{C}^9+H_{D}^9)$.
\begin{enumerate}
        \item  The $H_{X}^9$'s are bounded relative to the original Hamiltonian by $\gamma_9=0.02066720$.
        By computing the overlap $\eta'=0.05060345$ we get the relative bound $\gamma_0=0.01752971$.
        \item Alternatively, by using Prop 5 we bound $H^9$ relative to the full projector with $\gamma_R=0.8655232$, getting $\gamma_0=0.01788794$.
\end{enumerate}
        
  All of the above five different values of $\gamma_0$ give respective lower bounds on the AKLT gap via $\Delta_{\rm lower}=\gamma_0(1-\tilde{z} \eta)$, where $\tilde{z}=12$ and $\eta=0.041310153882$ was obtained in Sec.~\ref{diamond}.
\end{enumerate}
        

\begin{thebibliography}{99}
\bibitem{AKLT1}
 I. Affleck, T. Kennedy, E. H. Lieb, and H. Tasaki, Rigorous results on valence-bond
ground states in antiferromagnets, Phys. Rev. Lett. {\bf 59}, 799 (1987).

\bibitem{AKLT2} I. Affleck, T. Kennedy, E. H. Lieb, and H. Tasaki, Valence Bond Ground States in
Isotropic Quantum Antiferromagnets, Comm. Math. Phys. {\bf 115}, 477 (1988).


\bibitem{Gu}
Z.-C. Gu, and X.-G. Wen, Tensor-entanglement-filtering renormalization approach and symmetry protected topological order Phys. Rev. B \textbf{80} 155131 (2009).



\bibitem{Pollmann}
 F. Pollmann, E. Berg, A. M. Turner, and M. Oshikawa, Symmetry protection of topological order in one-dimensional quantum spin systems, Phys. Rev. B {\bf 85}, 075125 (2012).
 

\bibitem{Chen}
X. Chen, Z.-C. Gu, Z.-X. Liu, and X.-G. Wen, Symmetry-Protected Topological Orders in Interacting Bosonic Systems, Science {\bf 338},
1604 (2012).

\bibitem{Haldane83}
F. D.  M. Haldane, Continuum dynamics of the 1-d Heisenberg antiferromagnet:
identification with the O(3) nonlinear sigma model, Phys. Lett. {\bf 93}, 464 (1983).
\bibitem{Haldane83b}
 F. D. M. Haldane, Nonlinear field theory of large-spin Heisenberg antiferromagnets:
semiclassically quantized solutions of the one-dimensional easy-axis Neel state,
Phys. Rev. Lett. {\bf 50}, 1153 (1983).

\bibitem{Knabe}
S. Knabe, Energy gaps and elementary excitations for certain VBS-quantum antiferromagnets, J. Stat. Phys. {\bf 52}, 627 (1988).
\bibitem{Fannes1992}
M. Fannes, B. Nachtergaele, R.F. Werner, Finitely Correlated States on Quantum
Spin Chains, Commun. Math. Phys. {\bf 144}, 443 (1992).

 \bibitem{Gross}
 D. Gross and J. Eisert, Novel Schemes for Measurement-Based Quantum Computation, Phys. Rev. Lett. {\bf 98}, 220503 (2007).
 
 \bibitem{Brennen}
 G. K. Brennen and A. Miyake, Measurement-Based Quantum Computer in the Gapped Ground State of a Two-Body Hamiltonian, Phys. Rev. Lett. {\bf 101}, 010502
 (2008).
 
 
 \bibitem{Wei11}
 T.-C. Wei, I. Affleck, and R. Raussendorf,
 Affleck-Kennedy-Lieb-Tasaki State on a Honeycomb Lattice is a Universal Quantum Computational Resource,
  Phys. Rev. Lett. {\bf 106}, 070501 (2011).
  
 \bibitem{Miyake} A. Miyake, Quantum computational capability of a two-dimensional valence bond solid phase,
Ann. Phys. (Leipzig) {\bf 326}, 1656 (2011).
 
 \bibitem{Wei2013}
 T.-C. Wei, Quantum computational universality of spin-3/2 Affleck-Kennedy-Lieb-Tasaki states beyond the honeycomb lattice, Phys. Rev. A 88, 062307 (2013)
 \bibitem{Wei2014}
T.-C. Wei, P. Haghnegahdar, and R. Raussendorf, Hybrid valence-bond states for
universal quantum computation, Phys. Rev. A {\bf 90}, 042333 (2014).

 \bibitem{Wei15}
T.-C. Wei and R. Raussendorf, Universal measurement-based quantum computation with spin-2 Affleck-Kennedy-Lieb-Tasaki states, Phys. Rev A {\bf 92}, 012310 (2015).

\bibitem{PomataWei2020}
N. Pomata and T.-C. Wei, Demonstrating the Affleck-Kennedy-Lieb-Tasaki spectral gap on 2D degree-3 lattices,  Phys. Rev. Lett. {\bf124}, 177203 (2020).
\bibitem{LemmSandvikWang2020}
M. Lemm, A. W. Sandvik, and L. Wang, Existence of a Spectral Gap in the Affleck-Kennedy-Lieb-Tasaki Model on the Hexagonal Lattice, Phys. Rev. Lett. {\bf 124}, 177204 (2020).


\bibitem{Abdul}
 H. Abdul-Rahman, M. Lemm, A. Lucia, B. Nachtergaele, and A. Young,  A Class of Two-Dimensional AKLT Models with a Gap, in Analytic Trends in Mathematical Physics, Houssam Abdul-Rahman, Robert Sims, Amanda Young (Eds), Contemporary Mathematics vol 741, pp 1-21 (2020), American Mathematical Society.
  
\bibitem{PomataWei2019}
N. Pomata and T.-C. Wei, AKLT models on decorated
square lattices are gapped, Phys. Rev. B {\bf100}, 094429
(2019).

\bibitem{Param}
 S. A. Parameswaran, S. L. Sondhi, and D. P. Arovas,
 Order and disorder in AKLT antiferromagnets in three
dimensions,
Phys. Rev. B {\bf 79}, 024408 (2009).

\bibitem{KLT}
T. Kennedy, E. H. Lieb, and H. Tasaki, A two-dimensional isotropic quantum antiferromagnet with unique disordered ground state, J. Stat. Phys. {\bf 53}, 383 (1988).
\bibitem{Numerical1}
 A. Garcia-Saez, V. Murg, and T.-C. Wei, Spectral gaps of
Affleck-Kennedy-Lieb-Tasaki Hamiltonians using tensor network methods, Phys. Rev. {\bf B} 88, 245118 (2013).
\bibitem{Numerical2}
L. Vanderstraeten, M. Mariën, F. Verstraete, and J. Haegeman,
Excitations and the tangent space of projected entangled-pair
states, Phys. Rev. B {\bf 92}, 201111(R) (2015).
\end{thebibliography}
\end{document}